\documentclass[prb,reprint]{revtex4-2}

\usepackage{amsmath}  
\usepackage{amsfonts} 
\usepackage{graphicx} 
\usepackage{graphics}
\usepackage[dvipsnames]{xcolor}
\UseRawInputEncoding

\begin{document}

\title{Silent-enhancement of multiple Raman modes via tuning optical properties of graphene nanostructures}

\author{Asli Gencaslan$^{\bf (1)}$}
\author{Taner Tarik Aytas$^{\bf (1)}$}
\author{Hira Asif$^{\bf (1)}$}
\author{Mehmet Emre Tasgin$^{\bf (2)}$}
\author{Ramazan Sahin$^{\bf (1)}$}\email{rsahin@itu.edu.tr}

\affiliation{${\bf (1)}$ {Department of Physics, Akdeniz University, 07058 Antalya, Turkey}}
\affiliation{${\bf (2)}$ {Institute of Nuclear Sciences, Hacettepe University, 06800 Ankara, Turkey}}

\date{\today}

\begin{abstract}
Raman scattering signal can be enhanced through localization of incident field into sub-wavelength hot-spots through plasmonic nanostructures (Surface-enhanced Raman scattering-SERS). Recently, further enhancement of SERS signal via quantum objects are proposed by \cite{Postaci2018} without increasing the hot-spot intensity (\textit{silent-enhancement}) where this suggestion prevents the modification of vibrational modes or the breakdown of molecules. The method utilizes path interference in the non-linear response of Stokes-shifted Raman modes. In this work, we extend this phenomenon to tune the spectral position of \textit{silent-enhancement} factor where the multiple vibrational modes can be detected with a better signal-to-noise ratio, simultaneously. This can be achieved in two different schemes by employing either (i) graphene structures with quantum emitters or (ii) replacing quantum emitters with graphene spherical nano-shell in \cite{Postaci2018}. In addition, the latter system is exactly solvable in the steady-state. These suggestions not only preserve conventional non-linear Raman processes but also provide flexibility to enhance (silently) multiple vibrational Raman modes due to the tunable optical properties of graphene.
\end{abstract}

\maketitle

\section{Introduction}

Surface plasmons (SP) mostly utilize the optical properties of conventional metals (such as Au and Ag) to direct and manipulate incident light beyond the diffraction limit \cite{Brongersma2010, Lal2007}. The spatial confinement of incident energy through the coupling of light to electrons into sub-wavelength structures causes field enhancement up to $10^6$ times known as hot-spots \cite{Maier2007}. This enhancement/localization occurs in the form of Surface Plasmon Polaritons (SPP) \cite{Tomas2009} or can exist as Localized Surface Plasmons (LSP) which enable linear and non-linear plasmonic applications such as optical nanoantennas \cite{Giannini2011}, ultra-thin optical detectors \cite{Shen2013}, plasmonic sensors \cite{Stewart2008}, and Surface-Enhanced Raman Scattering (SERS) \cite{Campion1998}.

Among the above applications, SERS is a vibrational spectroscopy technique where Raman signal is measured with the spatially confined intense electromagnetic field \cite{Fleischmann1974, Otto2005, Ru2011, Ding2017}. Previous studies clearly indicate that this technique can chemically resolve single-molecule \cite{Nie1997, PhysRevLett.78.1667} in addition to determining its chemical bond type. Moreover, the double resonance scheme yields a strong non-linear process where input and Stokes/anti-Stokes shifted output energies are overlapped with two plasmon modes \cite{double_res_chu} besides they can be enhanced further via double Fano resonances  \cite{WOS:000369745300001,WOS:000301406800094,WOS:000319947800037}. Fano resonance is analogous to Electromagnetic Induced Transparency (EIT) except for the external driving source \cite{scully_quantum_1997,Hajebifard:17}. Fano resonance \cite{Anker2008, Giannini2011} paves the way for controlling both linear \cite{PhysRevB.101.035416,SAHIN2020124431,sahin_single_molecule,sahin_graphene_splitting} and non-linear \cite{emre_second_harmonic_steady,Singh2016,Turkpence2014, Asif_2022} plasmonic processes \cite{postaci_chapter} in many applications. Besides, it can be acquired by the coupling of dark plasmon modes or quantum objects (quantum emitter (QE), quantum dot (QD) or molecule, etc.) with bright plasmon modes \cite{Yildiz2015, Turkpence2014, Singh2016} provided that one of the coupled modes has a longer lifetime.

SERS is widely used \cite{Campion1998} for a variety of applications. Therefore, the extra enhancement in the Raman signal is very crucial \cite{Sharma2012} not only for better sensitivity but also for reaching the best spatial resolution \cite{Richards2003}. However, as the intensity of hot-spot increases to acquire stronger Raman signal, the target molecule might be prone to modification of vibrational modes and breakdown of the molecules \cite{Schaffer2001}. Therefore, overcoming these limitations without increasing the hot-spot intensity suggested by \cite{Postaci2018} can be very beneficial for detecting very low-Raman signals. On the other hand, graphene-based tunable plasmonic structures have been gathering huge attention from the research community. Therefore, we extend the proposed silent-enhancement \cite{Postaci2018} to tunable silent-enhancement in this study by using graphene nano-shell. The suggested scheme is satisfied by two-different approaches; (i) replacing graphene spherical shell with QE or (ii) employing graphene nanostructure in a double resonance scheme as explained below.

Graphene displays high carrier mobility and a broadband absorption spectra from near-infrared to terahertz frequencies \cite{Rao2010, Li2008, Koppens2014, Mueller2010} which makes it an interesting material due to its tunable optical properties \cite{Fang2014}. Regardless of its geometry, graphene indicates LSP or SPP resonances only in the infrared regime \cite{Jablan2009}. Group-IV nitrides (TiN, HfN, and ZrN, etc.) known as refractory transition metal nitrides (RTMN) exhibite metallic properties in VIS and IR \cite{Kumar2021}. Therefore, we employ Titanium Nitride (TiN) in our study. Moreover, TiN owns a much higher melting point $\sim 2930^o$C ($\sim 1064^o$C for Au and $\sim 961.8^o$C for Ag), possesses higher chemical stability and their optical properties can be adjusted by stoichiometry \cite{WOS:000304166900004} in addition to higher quality factor in the infrared region \cite{Juneja2019} as compared to Au. Furthermore, their plasmonic resonances occur in the infrared regime in which graphene plasmon spectrally overlap with Group-IV plasmon resonances \cite{Guler2013}. Therefore, through the proposed study not only the tuning of the silent-enhancement of Raman signal can be possible but also RTMNs and graphene structures are joined together for quantum plasmonic applications.

In the next section, we first define the setup for the coupled plasmonic system and then describe the analytical models for the tunable silent-SERS enhancement through non-linear Fano resonances. We suggested two different schemes for spectral tuning, either (i) graphene plasmon couples to QE which is then proved by numerical time-evolution of the equation of motions, or (ii) graphene spherical nano-shell is replaced with QE in the setup of \cite{Postaci2018}. The second case yields exactly a solvable system (time-independent in the steady-state) as explained before the conclusion part.

\begin{figure}[htb]
\centering
\fbox{\includegraphics[width=0.95\linewidth]{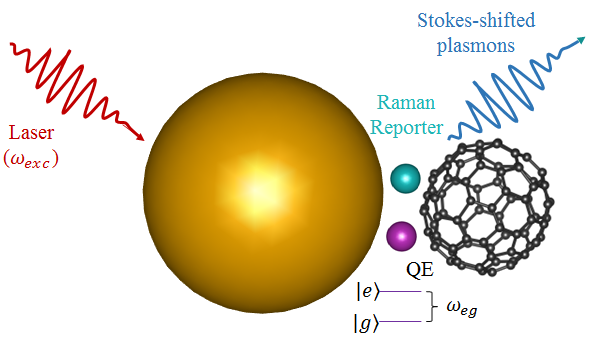}}
\caption{The QE (purple) placed between the gap of the dimer system consisting of TiN and graphene, and it interacts with the $\hat{a}_{G}$ ($\hat{a}_{R}$) mode where the Stokes-shifted signal is emerged. The Raman active molecule (turquoise) is also placed on the other side of the dimer gap. Thus, the signal received from the Raman active molecule is enhanced through QE without increasing the original hot-spot intensity.}
\label{fig:fig1}
\end{figure}

\section{Setup Configuration}

Here, we revisited the further enhancement in SERS signal with the double plasmon-resonance studied in \cite{Postaci2018} since our main motivation is to adjust spectrally the silent-enhancement factor explained in \cite{Postaci2018} as a proof-of-principle concept. The classical system consists of plasmonic structures exhibiting two-plasmon modes (overlapping with the wavelength of laser source and Stokes-shifted plasmon mode), and a Raman reporter molecule. In addition, a quantum emitter (QE) is also placed very close to hot-spot for non-linear Fano enhancement as shown in Fig. \ref{fig:fig1}.

\begin{figure}[htbp]
\centering
\fbox{\includegraphics[width=0.95\linewidth]{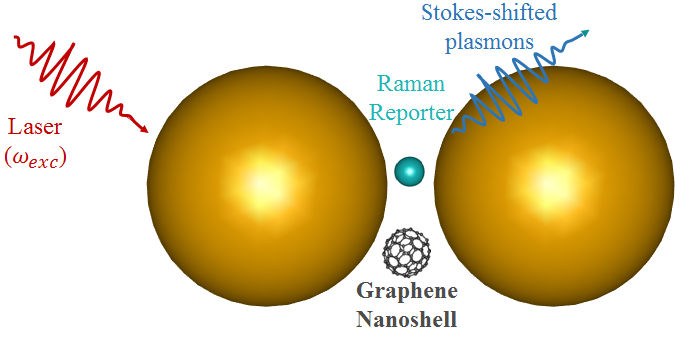}}
\caption{The dimer system consists of two TiN nanoparticles. For spectral tuning of Raman modes graphene is replaced with QE for further-enhancement.}
\label{fig:fig2}
\end{figure}

The graphene nano-shell is replaced with one of the metal nanoparticles in our setup. The Raman process and silent-enhancement occur in such a way that, an incident laser ($\omega_{exc}$) excites one of the plasmon modes ($\hat{a}$) and the Raman-active molecule positioned near the hot-spot induces the Stokes-Raman process. The Stokes-shifted Raman plasmon ($\hat{a}_R$) has lower energy and its spectral position depends on the vibrational mode of the Raman-active molecule. Due to the coupling between QE and Stokes-shifted plasmon mode, the SERS signal can be enhanced further. In addition, this further enhancement can be tuned in two different ways by adjusting (i) the spectral position of the second plasmon mode (Fig. \ref{fig:fig1}) or (ii) by replacing QE with graphene nanostructure since its optical properties can be tuned by an external electric field (Fig. \ref{fig:fig2}).

Since the second approach employs graphene nanostructure instead of QE, the dynamics of the coupled system can be derived from the exactly solvable (time-independent) equation of motions in the steady-state as explained in the next section.

\section{Theoretical Model and Numerical Calculations}

In this part, we introduce the analytical models. We first derive the total Hamiltonian of the non-linear system composed of two plasmon modes, laser source, Raman conversion process, molecular vibrations, and energy of QE or graphene. From these energies, equations of motion can be obtained by solving the Heisenberg equation. Then, we acquired steady-state amplitudes for the Stokes-shifted Raman signal which can be enhanced and tuned without disturbing the plasmonic system, and hence we are able not only to enhance it further but also to adjust it without disturbing the plasmonic system.

\subsection{Tuning the silent-enhancement with QE}

In the first approach, we consider a plasmonic dimer consisting of TiN nano-particle and graphene nano-shell separated by a distance of $6$ nm from each other with diameters of $200$ nm and $10$ nm, respectively. The laser source ($\omega_{\rm exc}$) excites the LSP mode of the TiN nano-particle. The non-linear Raman conversion process (the Raman active molecule is responsible for this) creates Stokes-shifted plasmon mode of graphene nano-shell with frequency $\omega_{R}$. Graphene plasmons strongly interact with QE, considered as a two-level system with ground state $| g \rangle $  and excited state $| e \rangle $ and the corresponding energy is $\omega_{eg}$. The total Hamiltonian of the system (summation of each term in Eq. \ref{eq:Eq1}) can be explained as follows using the Jaynes-Cumming model \cite{Jaynes1963};

\begin{subequations}
\begin{align}
\hat{H}_{0}&= \hbar \Omega \hat{a}^\dagger \hat{a}+\hbar \Omega_{G} \hat{a}_{G}^\dagger \hat{a}_{G}+\hbar \Omega_{ ph} \hat{a}_{ ph} ^\dagger \hat{a}_{ ph} \\
\hat{H}_{P}&= i\hbar \varepsilon  \hat{a}^\dagger e^{-i\omega_{exc} t} +\textit{{ h.c}}\\
\hat{H}_{QE}&= \hbar \omega_{eg}| e \rangle \langle e|\\
\hat{H}_{ int}&= \hbar f \hat{a}_{G}^\dagger | g \rangle \langle e| +\textit{h.c}\\
\hat{H}_{R}&=   \hbar \chi ( \hat{a}_{G} ^\dagger \hat{a}_{ph} ^\dagger \hat{a} +\textit{{ h.c}}  )+ ( i\hbar \varepsilon_ {ph} \hat{a}_{ph}^\dagger e^{-i\omega_{ ph } t} +\textit{{ h.c}})
\end{align}
\label{eq:Eq1}
\end{subequations}

$\hat{H}_0$ includes not only energies of plasmon modes but also energies of phonons, where $\hat{a}^\dagger(\hat{a})$, $\hat{a} _G^\dagger(\hat{a}_G)$, and $\hat{a} _{ph}^\dagger(\hat{a}_{ph})$ are creation (annihilation) operators of excited plasmon mode, Stokes-shifted plasmon mode (graphene plasmon mode), and phonon mode respectively. Moreover, $\Omega$, $\Omega_G$, and $\Omega_{ph}$  are the resonance frequencies of corresponding modes. The coupling strength ($f$) between Stokes-shifted plasmon mode and QE is added as $\hat{H}_{int}$ in the total Hamiltonian and $\hat{H}_{P}$ is the energy of laser ($\omega_{\rm exc}$). Complex amplitudes (Eq. \ref{eq:Eq3}) ${\alpha}$, $ {\alpha}_{G} $, $ {\alpha}_{ph} $, ${\rho}_{ ge}$, and ${\rho}_{ ee}$ are calculated \cite{Premaratne2017} from Heisenberg equations ($i \hbar \dot{\hat{a}}=[\hat{a},\hat{H}]$) by replacing with the operators $ \hat{a} $, $ \hat{a}_{G}$,  $ \hat{a}_{ph}$,  $ \hat{\rho}_{ge}$, and $\hat{\rho}_{ee}$ (the equations of motion are given below, Eq. \ref{eq:Eq2}).

\begin{subequations}
\begin{align}
\dot{\alpha} &= -i \Omega{\alpha}-i\chi^{*}{\alpha}_{ ph}{\alpha}_{G} + \varepsilon e^{-i\omega_{exc} t} \label{2a}\\
\dot{\alpha}_{ ph}&=   -i \Omega_{ ph}{\alpha}_{ ph}-i\chi{\alpha}_{ G}^{\ast}  {\alpha}+ \varepsilon_{ ph} e^{-i\omega_{ ph} t} \label{2b}\\
\dot{\rho}_{ ge} &= -i \omega_{ eg} {\rho}_{ ge} + i f^\ast y {\alpha}_{ G} \label{2c}\\
\dot{\rho}_{ ee} &=  i f \alpha_{G}^{\ast}  {\rho}_{ge} - i f^\ast \alpha_{G}  {\rho}_{ge} ^{\ast}  \label{2d}\\
\dot{\alpha}_{G} &=  -i \Omega_{G}{\alpha}_{G}-if {\rho}_{ge} -i\chi{a}_{ ph}^{\ast}  \alpha \label{2e}.
\end{align}
\label{eq:Eq2}
\end{subequations}

$\gamma$, $\gamma_G$, $\gamma_{ph}$ , $\gamma_{ge}$, and $\gamma_{ee}$ are the decay rates of excited and Stokes-shifted plasmon modes, vibrational mode and graphene plasmons, and density matrix elements, respectively. In the steady-state, the system acts as interacting oscillators with the driving frequency of $\omega_{exc}$. Therefore, the set of equations of motion becomes as follows.

\begin{subequations}
\begin{align}
&[i \left(\Omega - \omega_{exc} \right) + \gamma ] \alpha =\varepsilon - i\chi^{\ast} \alpha_{ph} \alpha_{G} \\
&[i \left(\Omega_{ph} - \omega_{ph} \right) + \gamma_{ph} ] \alpha_{ph} = - i\chi \alpha_{G}^\ast \alpha + \varepsilon_{ph} \\
&[i \left(\omega_{eg} - \omega_{R} \right) + \gamma_{ge} ] \rho_{ge} = if^\ast y \alpha_{G} \\
&\gamma_{ee} \rho_{ee} = if \alpha_{G}^{\ast} \rho_{ge} - if^\ast \alpha_{G} \rho_{ge}^{\ast}\\
&[i \left(\Omega_{G} - \omega_{R} \right) + \gamma_{G} ] \alpha_{G} = -i f \rho_{ge} - i\chi \alpha_{ph}^{\ast} \alpha
\end{align}
\label{eq:Eq3}
\end{subequations}

In the steady-state, all modes oscillate in the form of Eq. \ref{eq:Eq4}.

\begin{equation}
\begin{aligned}
    \alpha (t)&=\tilde{\alpha}e^{-i \omega_{exc} t}\\
    \alpha _{ph} (t)&=\tilde{\alpha}_{ph}e^{-i \omega_{ph} t}\\
    \rho_{ge} (t)&=\tilde{\rho}_{ge}e^{-i \omega_{R} t}\\
    \rho_{ee} (t)&=\tilde{\rho}_{ge}\\
    \alpha _{G} (t)&=\tilde{\alpha}_{G}e^{-i \omega_{R} t}\\
\end{aligned}
\label{eq:Eq4}
\end{equation}

Due to the conservation of energy in the Raman processes and cancellation of exponential terms, the steady-state amplitude of Stokes-shifted Raman conversion is obtained (Eq. \ref{eq:Eq5}),

\begin{eqnarray}
\tilde{\alpha}_R={\frac{- i \chi {\varepsilon_{ph}}^\ast}{{\beta_{ph}}^\ast {f}_{1}(\omega_{eg})- |\chi|^2 |\tilde{\alpha}|^2
}}{\tilde{\alpha}} \: ,
\label{eq:Eq5}
\end{eqnarray}

where

\begin{equation}
 {f}_{1}(\omega_{eg}) = [i(\Omega_G -\omega_{\rm R})+\gamma_G] - \frac{|f|^2 y} {[i(\omega_{eg} - \omega_{R})+\gamma_{eg}]},
\label{eq:Eq6}
 \end{equation}

and

\begin{equation}
\beta_{ph}=i (\Omega_{ph} - \omega_{ ph})+\gamma_{ph}.
\label{eq:Eq7}
\end{equation}

Here, $y$=$\rho_{ee}-\rho_{gg}$ is the population inversion parameter. Steady-state amplitude of Stokes-shifted plasmon mode (Eq. \ref{eq:Eq5}) clearly indicates that coupling yields extra enhancement in $\tilde{\alpha}_R$ by minimizing the denominator even without changing the hot-spot intensity ($\tilde{\alpha}$). Through time-evolution of Eq. \ref{eq:Eq5}, one can obtain the tunable silent-enhancement of SERS mode as detailed below.

\begin{figure}[htbp]
\centering
\includegraphics[width=0.47\textwidth]{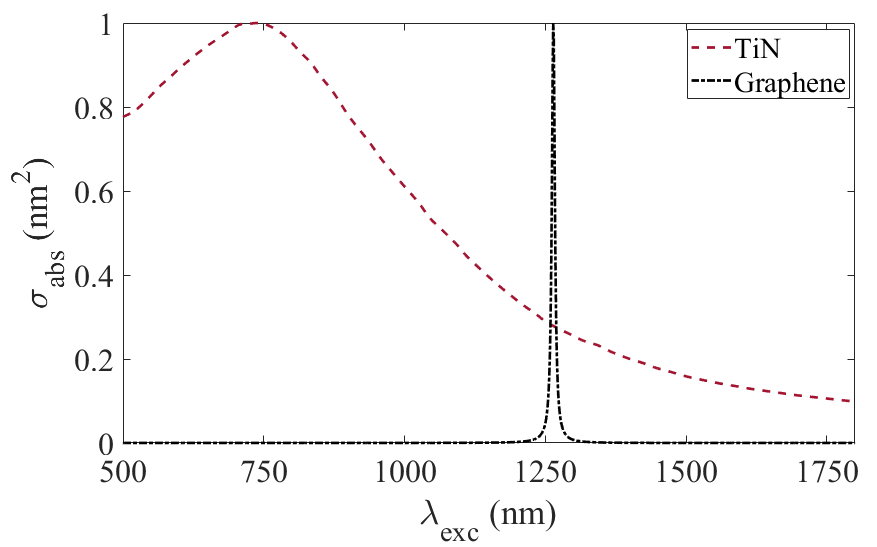}
\caption{Absorption cross-sections ($\sigma_{abs}$) of TiN nanoparticle (diameter is $200$ nm) and graphene nano-shell (diameter is $10$ nm). The thickness of the graphene is taken as $0.5$ nm. For the dielectric constants of TiN and graphene, \cite{Naik2011} and \cite{Vakil2011} are used, respectively.}
\label{fig:fig3}
\end{figure}

\begin{figure}[htbp]
\centering
\includegraphics[width=0.47\textwidth]{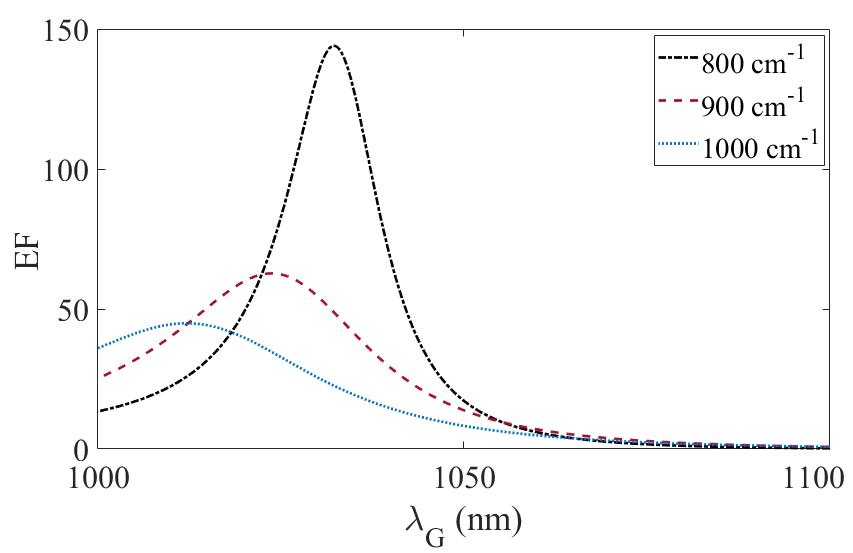}
\caption{Enhancement factor of different Stokes-shifted Raman modes ($800$ $cm^{-1}$, $900$ $cm^{-1}$, $1000$ $cm^{-1}$) at the fixed level-spacing of QE $\lambda_{eg}$=$1281$ nm  with coupling strength $f$=$0.1\omega_{exc}$.}
\label{fig:fig4}
\end{figure}

\begin{figure}[htbp]
\centering
\includegraphics[width=0.47\textwidth]{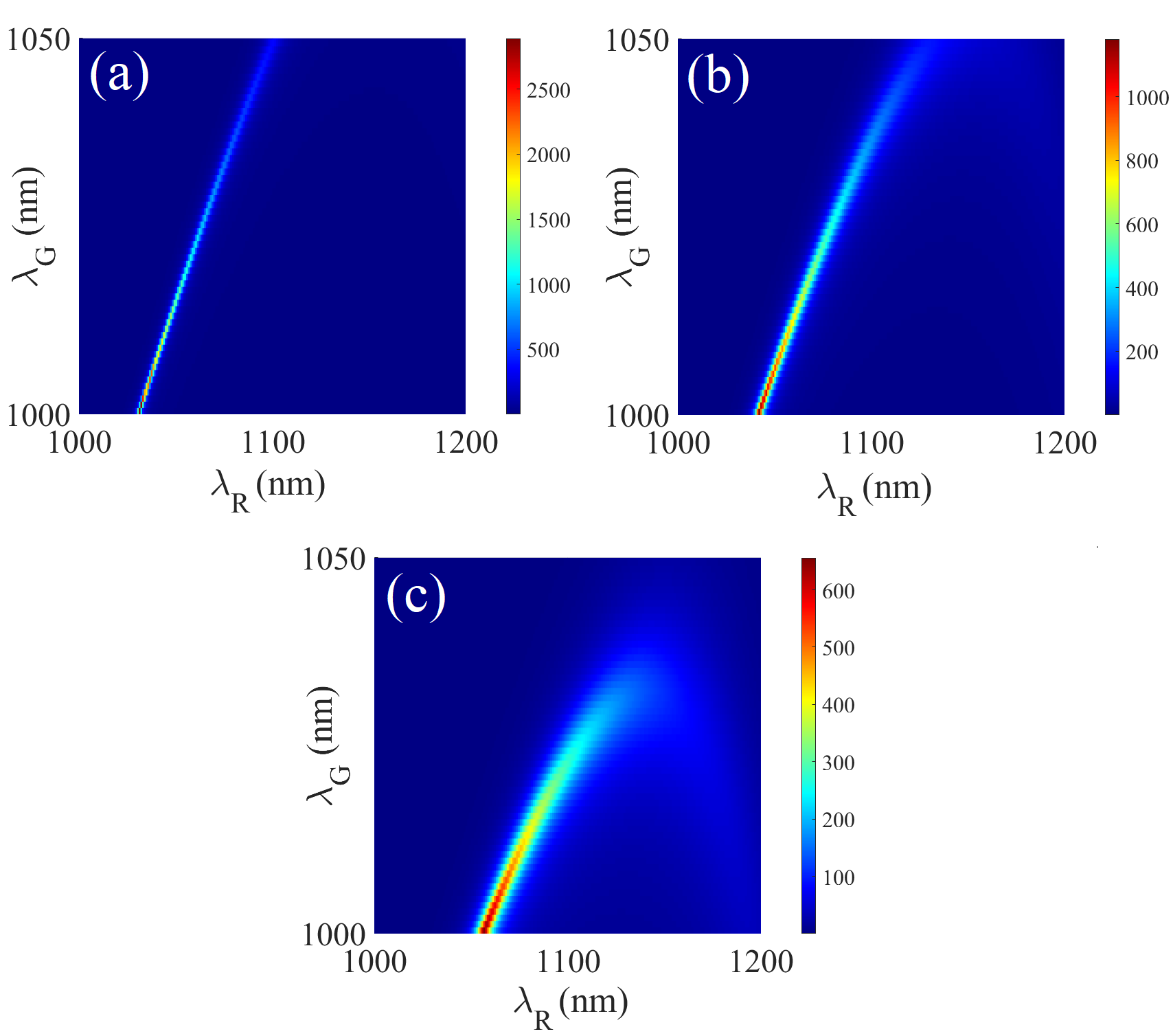}
\caption{Enhancement factor with respect to wavelength of graphene plasmon resonance and Stokes-shifted Raman mode for (a) $f$=$0.08 \omega_{exc}$, (b) $f$=$0.09\omega_{exc}$, (c) $f$=$0.1\omega_{exc}$. We kept $\lambda_{eg}$=$1281$ nm constant in these calculations.}
\label{fig:fig5}
\end{figure}

We first determine the specific values for the resonance wavelengths of plasmon modes through a 3D solution of the Maxwell equation based on Boundary Element Method (BEM) with a toolbox of MNPBEM \cite{HOHENESTER2012370} in Matlab. Absorption profiles of TiN nano-particle and graphene nano-shell are shown in Fig. \ref{fig:fig3}. From Fig. \ref{fig:fig3}, LSP resonances of TiN nano-particle and graphene nano-shell are found as $740$ nm and $1264$ nm, respectively. In addition, the decay rates of TiN and graphene are taken as $\sim 10^{13}$  Hz and $\sim 10^{12}$ Hz. Moreover, the Lorentzian dielectric constant \cite{Wu2010} is employed for the dielectric constant of an auxiliary molecule (QE). We take its decay rate as $\gamma_{eg}\sim10^{13}$ Hz, throughout the numerical calculations ($\gamma_{ph}=10^{-3}\omega_{exc}$, $\chi=10^{-5}\omega_{exc}$, $\varepsilon_{ph}=\varepsilon=10^{-1}\omega_{exc}
$).

We utilized the above parameters into the Eq. \ref{eq:Eq5}. Furthermore, the time evolution of Eq. \ref{eq:Eq2} and Eq. \ref{eq:Eq5} are performed which give us the amplitude of Stokes-shifted Raman signals through silent-enhancement as shown in Fig. \ref{fig:fig4}. The enhancement factor presented in this manuscript is found by dividing the intensity of the Stokes-shifted Raman signal in the presence of a QE/graphene spherical shell (which means that the coupling strength (f) is not zero) by the intensity of the Stokes-shifted Raman signal without the QE/graphene spherical shell ($f$=0). As we suggested in our study, tuning the plasmon resonance of graphene structure provides enhancement (at least 2-orders of magnitude) in a variety of Stoke-shifted Raman modes ($800$  $cm^{-1}$, $900$ $cm^{-1}$, and $1000$ $cm^{-1}$)

In order to explore the effect of coupling strength between the graphene plasmons modes and QE on the ability to tune the enhancement factor, we repeated the same calculation for a variety of $f$ ($f$=$0.08\omega_{exc}$, $f$=$0.09\omega_{exc}$, $f$=$0.1\omega_{exc}$) values. Fig. \ref{fig:fig5} shows the extra enhancement of Raman signal. It can be clearly seen from Fig. \ref{fig:fig5} that changing the plasmon resonance of graphene structure provides multiple enhancements of different Raman modes. As we have mentioned in the previous section, this can be straightforward by applying a potential difference to graphene structures.

\subsection{Adjusting silent-enhancement without QE}

In this context, the dimer system is composed of two TiN nanoparticles with diameters of $400$ nm and $200$ nm separated by $12$ nm from each other, as shown in Fig. \ref{fig:fig2}. The graphene nano-shell is positioned very close to hot-spot between the TiN nanoparticles. The Raman active molecule is brought to the other side of the dimer gap so that non-linear Raman conversion can occur. In this scheme, graphene nano-shell acts as if it is QE since it has a strong absorption profile (see Fig. \ref{fig:fig3}) and its plasmon-lifetime is longer than metal plasmons. The effective Hamiltonian of such a system in this case can be written as the sum of its components (Eq. \ref{eq:Eq8});

\begin{subequations}
\begin{align}
\hat{H}_{ 0}&= \hbar \Omega \hat{a}^\dagger \hat{a}+\hbar \Omega_{ R} \hat{a}_{ R}^\dagger \hat{a}_{ R}+\hbar \Omega_{ ph} \hat{a}_{ ph} ^\dagger \hat{a}_{ ph} \\
\hat{H}_{ P}&= i\hbar \varepsilon  \hat{a}^\dagger e^{-i\omega_{exc} t} +\textit{{ h.c}}\\
\hat{H}_{G}&= \hbar \Omega_{ G} \hat{a}_{ G}^\dagger \hat{a}_{ G}\\
\hat{H}_{ int}&=  \hbar f \hat{a}_{ R}^\dagger \hat{a}_{ G} +\textit{{ h.c}}\\
\hat{H}_{ R}&= ( \hbar \chi \hat{a}_{ R} ^\dagger \hat{a}_{ ph} ^\dagger \hat{a} +\textit{{ h.c}}) + (i\hbar \varepsilon_ {ph} \hat{a}_{ph}^\dagger e^{-i\omega_{ _{ph} }t} +\textit{{ h.c}}).
\end{align}
\label{eq:Eq8}
\end{subequations}

Here, apart from the previous case, $\hat{H}_{G}$ is the energy of the graphene nano-shell and it is responsible for the non-linear Fano enhancement of the SERS signal. $\hat{H}_{int}$ represents the interaction between Stokes-shifted plasmon mode and LSP of graphene nano-shell where $\hat{H}_{ R}$ emerges energy conservation in this non-linear Raman process. Similarly, the equations of the motion (Eq. \ref{eq:Eq9}) can be obtained from the Heisenberg equations

\begin{subequations}
\begin{align}
\dot{\hat{a}} &=   -i \Omega\hat{a}-i\chi^\ast\hat{a}_{ ph}\hat{a}_{R} + \varepsilon e^{-i\omega_{exc} t} \label{9a}\\
\dot{\hat{a}}_{ph}&=   -i \Omega_{ ph}\hat{a}_{ ph}-i\chi\hat{a}_{ R}^\dagger \hat{a}+ \varepsilon_{ ph} e^{-i\omega_{ ph} t} \label{9b}\\
\dot{\hat{a}}_{ G}&=   -i \Omega_{ G}\hat{a}_{ G}-if^\ast\hat{a}_{R}  \label{9c}\\
\dot{\hat{a}}_{ R}&=   -i \Omega_{ R}\hat{a}_{ R}-i\chi\hat{a}_{ ph}^\dagger \hat{a} -if\hat{a}_{ G}  \label{9d}
\end{align}
\label{eq:Eq9}
\end{subequations}

We measure the intensity of plasmon modes or scattered fields in SERS experiments. Therefore, the operators $ \hat{a} $, $ \hat{a}_R$, $ \hat{a}_{ph}$ and $ \hat{a}_{G}$  are replaced with complex numbers $ {\alpha} $, $ {\alpha}_R $, $ {\alpha}_{ph} $ and $ {\alpha}_{G} $, respectively, as follows (Eq. \ref{eq:Eq10}).

\begin{subequations}
\begin{align}
\dot{{\alpha}} &=  (-i \Omega-\gamma){\alpha}-i {{\chi}^\ast} {\alpha_{ph}} {\alpha_R}+\varepsilon e^{-i\omega_{exc} t} \\
\dot{{\alpha}}_{ph} &=  (-i \Omega_{ph}-\gamma_{ph}){\alpha_{ph}}- i \chi {{\alpha_R}}^\ast {\alpha}+\varepsilon_{ph} e^{-i\omega_{\rm ph} t} \\
\dot{{\alpha}}_{G} &=  (-i \Omega_{G}-\gamma_{G}){\alpha_{G}}- i {f}^\ast {\alpha_R} \\
\dot{{\alpha}}_{R} &=  (-i \Omega_R -\gamma_R) {\alpha_R}-i \chi {{\alpha_{ph}}}^\ast {\alpha}- i f {\alpha_{G}}.
\end{align}
\label{eq:Eq10}
\end{subequations}

In the steady-state, complex amplitudes in Eq. \ref{eq:Eq10} yield the following equations;

\begin{equation}
\begin{aligned}
    \alpha (t)&=\tilde{\alpha}e^{-i \omega_{exc} t}\\
    \alpha _{ph} (t)&=\tilde{\alpha}_{ph}e^{-i \omega_{ph} t}\\
    \alpha _{G} (t)&=\tilde{\alpha}_{G}e^{-i \omega_{R} t}\\
    \alpha _{R} (t)&=\tilde{\alpha}_{ph}e^{-i \omega_{R} t}.\\
\end{aligned}
\label{eq:Eq11}
\end{equation}

Putting the complex amplitudes (Eq. \ref{eq:Eq11}) into the Eq. \ref{eq:Eq10} and cancellation of exponential terms yield Eq. \ref{eq:Eq12}.

\begin{subequations}
\begin{align}
[i (\Omega - \omega_{exc})+\gamma)]\tilde{\alpha} &= \varepsilon - i {{\chi}^\ast}{ \tilde{\alpha}_{ph}}{ \tilde{\alpha}_{R}} \label{12a} \\
[i (\Omega_{ph} - \omega_{ ph})+\gamma_{ph}]\tilde{\alpha}_{ph} &= - i \chi {{\tilde{\alpha_{R}}}}^\ast \tilde{\alpha} + \varepsilon_{ph} \label{12b}\\
[i (\Omega_{G} - \omega_{ R})+\gamma_{G}]\tilde{\alpha} &= - i {f}^\ast {\tilde{\alpha}_{R}} \label{12c}\\
[i (\Omega_{R} - \omega_{ R})+\gamma_{R}]\tilde{\alpha}_{R} &= - i \chi {\tilde{\alpha}_{ph}}^\ast {\tilde{\alpha}} - i f {\tilde{\alpha}_{G}} \label{12d}
\end{align}
\label{eq:Eq12}
\end{subequations}

$\gamma$, $\gamma_{ph}$ , $ \gamma_{G} $, and  $\gamma_R$ are the decay rates of $ \hat{a} $, $ \hat{a}_{ph}$, $ \hat{a}_{G}$, and $ \hat{a}_{R}$. Finally, one can obtain the steady-state amplitude for Stokes-shifted Raman mode ($\tilde{\alpha}_R$) as follows (Eq. \ref{eq:Eq13});

\begin{equation}
\tilde{\alpha}_R={\frac{- i \chi {\varepsilon_{ph}}^\ast}{{\beta_{ph}}^\ast {f}_{2}(\Omega_{G})- |\chi|^2 |\tilde{\alpha}|^2
}}{\tilde{\alpha}},
\label{eq:Eq13}
\end{equation}

\begin{equation}
 {f}_{2}(\Omega_{G}) = [i(\Omega_R -\omega_{\rm R})+\gamma_R] + \frac{|f|^2} {[i(\Omega_{G} - \omega_{\rm R})+\gamma_{G}]},
\label{eq:Eq14}
\end{equation}

and

\begin{equation}
\beta_{ph}=i (\Omega_{ph} - \omega_{ ph})+\gamma_{ph}.
\label{eq:Eq15}
\end{equation}

Eq. \ref{eq:Eq13} resembles very much to Eq. \ref{eq:Eq5} except for $'y'$, '\textit{population inversion}'. The absence of this parameter makes this system exactly solvable (time-evolution of the system is not necessary). Similarly, one can enhance the $\tilde{\alpha}_R$ via minimizing the denominator of the Eq. \ref{eq:Eq13} without changing the $\tilde{\alpha}$.

\begin{figure}[htbp]
\centering
\includegraphics[width=0.47\textwidth]{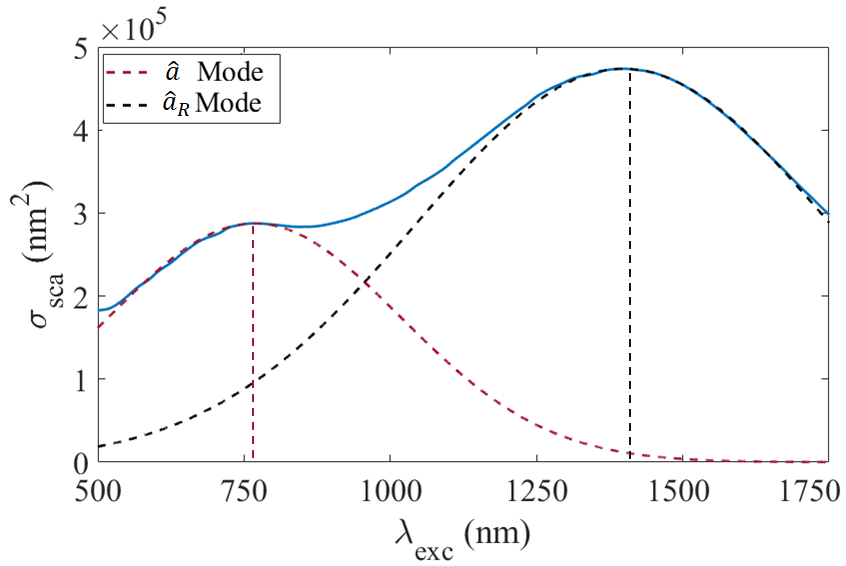}
\caption{Scattering cross-section of TiNs as a function of wavelength. Each profile of the double resonance scheme is individually Gaussian fit. Red dashed lines represent the first plasmon mode of the dimer, while black dashed lines represent the stock-shifted Raman plasmon mode.}
\label{fig:fig6}
\end{figure}

\begin{figure}[htbp]
\centering
\includegraphics[width=0.47\textwidth]{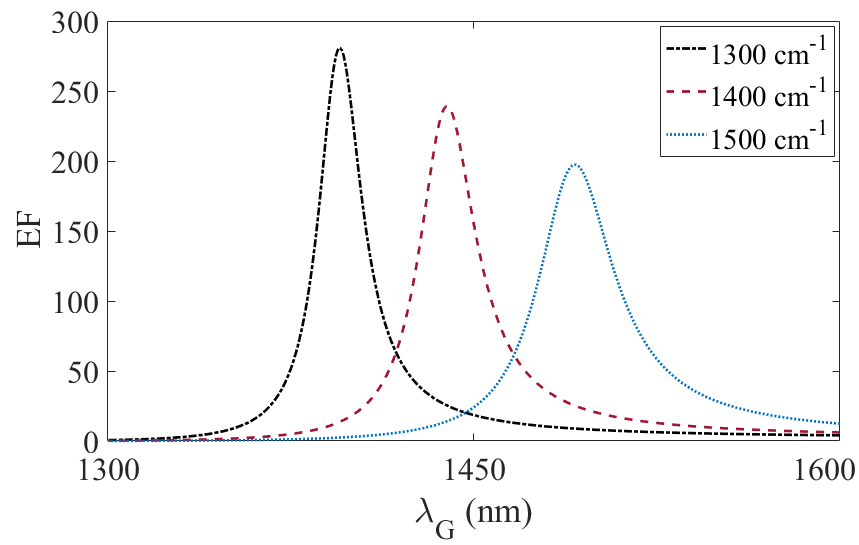}
\caption{The enhancement factor is calculated for different Stokes-shifted Raman modes ($1300$ $cm^{-1}$, $1400$ $cm^{-1}$, $1500$ $cm^{-1}$) as a function of excitation wavelength for graphene with coupling strength taken as $f$=$0.1\omega_{exc}$.}
\label{fig:fig7}
\end{figure}

In this part, we repeated the same procedure to clearly indicate the tuning capability of an extra enhancement of the Raman signal. We first determine the properties of plasmon modes of TiN nanoparticles (dimer). The solution of 3D Maxwell equations through MNPBEM calculations is shown in Fig. \ref{fig:fig6} for TiN dimer. The dimer system indicates two-plasmon modes at $766$ nm and $1397$ nm. And, we take their decay rates as $\sim 10^{13}$ Hz.

Since this system provides exactly solvable non-linear process in the steady-state, the time-evolution of Eq. \ref{eq:Eq13} is not required. We assume again that the wavelength of the excitation laser is $1064$ nm. At these conditions, numerical calculations of Eq. \ref{eq:Eq13} clearly indicate enhancement of different Raman modes ($1300$ $cm^{-1}$, $1400$ $cm^{-1}$, $1500$  $cm^{-1}$) as shown in Fig. \ref{fig:fig7}.

\begin{figure}
\centering
\includegraphics[width=0.47\textwidth]{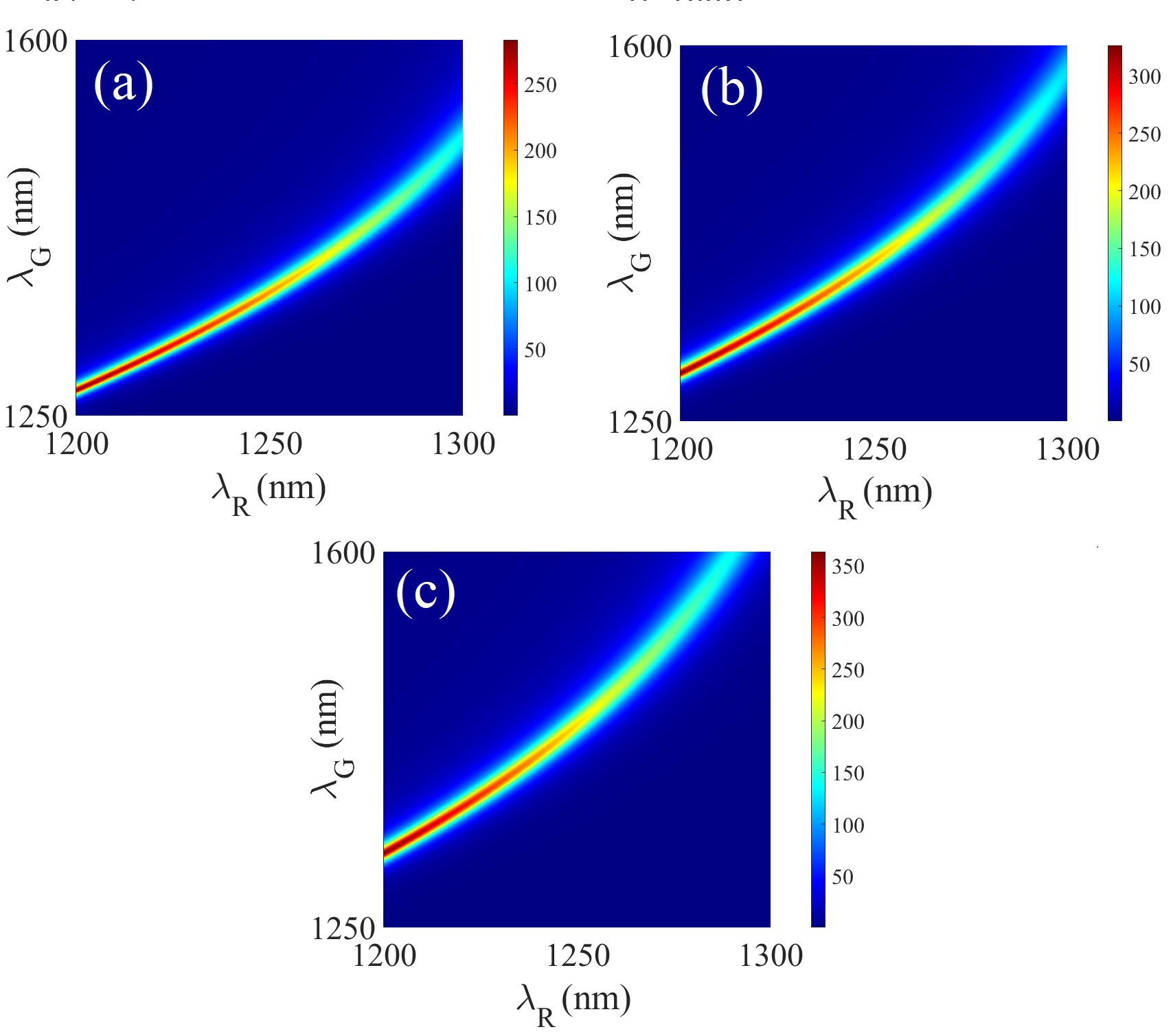}
\caption{Enhancement factor with respect to wavelength of graphene plasmon resonance and Stokes-shifted Raman mode for (a) $f$=$0.08 \omega_{exc}$, (b) $f$=$0.09\omega_{exc}$, (c) $f$=$0.1\omega_{exc}$.}
\label{fig:fig8}
\end{figure}

We also analyzed the effect of coupling strength between second plasmon mode of TiN dimer and graphene nano-shell on the ability to tune enhancement factor for a variety of $f$ ($f$=$0.08\omega_{exc}$, $f$=$0.09\omega_{exc}$, $f$=$0.1\omega_{exc}$) values. Fig. \ref{fig:fig8} clearly indicates that an extra enhancement of Raman signal can be adjusted provided that the spectral position of LSP resonance of graphene nano-shell can be tuned.

\section{Conclusion}
Due to coupling between Stokes-shifted plasmon mode and (i) QE or (ii) graphene nano-shell, we propose adjustment of a silent-enhancement factor in SERS experiments. Since our analysis indicate both silent-enhancement of SERS signal and adjusting its spectral position with graphene nano-shell structures, we disregard the coupling of QE or graphene nano-shell to the excited plasmon mode. Moreover, we propose the involvement of RTMN materials first time to the best of our knowledge in quantum plasmonic applications. We join the unique properties of graphene, RTMN materials, and non-linear Fano resonances together in this study. Therefore, our suggestions can be very beneficial for increasing the amplitude of the Raman signal of different Stokes-shifted Raman modes as the multi-mode Raman measurements are required.

\begin{acknowledgments}

R.S., M.E.T, A.G. and T.T.A. acknowledge support from TUBITAK-Project No. 121F030.

\end{acknowledgments}


\end{document}